\documentclass[12pt]{article}
\usepackage{graphicx}
\begin{document}
\begin{center}

{\bf Cherenkov gluons\\ (predictions and proposals)}\footnote{The talk
at the seminar during the CERN TH-workshop "Heavy Ion Collisions at the LHC. 
Last Call for Predictions." May 14th - June 8th 2007.}\\
\bigskip

I.M. Dremin\footnote{email: dremin@lpi.ru} 

{\it Lebedev Physical Institute, Moscow, Russia}\\

\end{center}
\begin{abstract}
The coherent hadron production analogous to Cherenkov radiation of photons 
gives rise to the ring-like events. Being projected on the ring diameter 
they produce the two-bump structure recently observed for the away-side 
jets at RHIC. The position of the peaks and their height determine such
properties of the hadronic medium as its nuclear refractive index, the 
parton density, the free path length and the energy loss of Cherenkov gluons.
Cherenkov gluons may be responsible for the asymmetry of dilepton mass
spectra near $\rho $-meson observed in experiment. Beside comparatively low 
energy gluons observed at RHIC, there could be high energy gluons at LHC, 
related to the high energy region of positive real part of the forward 
scattering amplitude and possessing different characteristics. This would 
allow to scan ($x, Q^2$)-plane determining the parton densities in its
various regions.

\end{abstract}

\section{Introduction.}

Analogous to Cherenkov photons \cite{cher, tf, tamm}, the Cherenkov gluons 
\cite{1, 2, 3, 4, a} can be emitted in hadronic collisions.
Considered first for processes at very high energies \cite{1, 2}, the idea about 
Cherenkov gluons was extended to resonance production \cite{3, 6, 7}. 

For Cherenkov effects to be pronounced in ordinary or nuclear matter, the 
(either electromagnetic or nuclear) refractive index of the medium $n$ should 
be larger than 1. There exists the general relation (see, e.g., \cite{8}) 
between the refractive index and the forward scattering amplitude $F(E, 0^o)$:
\begin{equation}
\Delta n={\rm Re}n-1 = \frac{8\pi N_s {\rm Re}F(E, 0^o)}{E^2}.    \label{delt}
\end{equation}
Here $E$ is the photon (gluon) energy, $N_s$ is the density of the scattering 
centers in the medium. We shall use this relation as a starting point for all 
results on Cherenkov gluons. The necessary condition for Cherenkov radiation is
\begin{equation}
\Delta n >0  \;\;\;\; {\rm or} \;\;\;\; {\rm Re}F(E, 0^o) >0.    \label{ref}
\end{equation}
If these inequalities are satisfied, Cherenkov photons (gluons) are emitted 
along the cone with half-angle $\theta _c$ in the {\bf rest} system of the 
{\bf infinite} medium determined by $n$:
\begin{equation}
\cos \theta_c=\frac {1}{\beta n},   \label{cos}
\end{equation}
where $\beta (\approx 1$ for relativistic partons) is the ratio of the 
velocities of the parton and (in-vacuum) light. 

Let us stress that the notion of the angle is not relativistically invariant.
Therefore one should be careful in choosing the coordinate system where to
apply Eq. (\ref{cos}) in nucleus-nucleus collisions. This problem is discussed
in more detail at the end of the paper.

In classical electrodynamics, it is the dipole excitation and polarization 
of atoms by the electromagnetic field of charged beams moving in the medium 
which results in the Breit-Wigner shape of the amplitude $F(E, 0^o)$. The 
relation (\ref{delt}) demonstrates that the forward scattering amplitudes 
account in some way for properties of excited states as well. Atoms behaving as 
oscillators provide the refractive indices larger than 1 within their 
low-energy wings (see, e.g., Fig. 31-5 in vol. 1 of Feynman lectures \cite{9}). 
The target rest system is precisely defined. In distinction to Bremsstrahlung
where transverse fields are important, the Cherenkov radiation is induced by 
the longitudinal component of the polarization vector.

The forward scattering amplitude for gluons is infinite. To define the 
refractive index in the absence of the theory of nuclear media (for a simplified 
approach see \cite{a}) I prefer to rely on our knowledge about hadronic 
reactions translated in partonic language to develop phenomenological models.

There are two common features for all hadron-hadron collisions. First, the
prominent resonances are formed at rather low energies. They are described by 
the Breit-Wigner amplitudes which have the positive real part in their low-mass
wings. In hadronic medium, there should be some modes (quarks, gluons or their 
preconfined bound states, condensates, blobs of hot matter...?) which can get 
excited by the gluon field of the impinging parton and radiate coherently if 
$n>1$. The necessary condition (\ref{ref}) for the Cherenkov effect is 
satisfied for those gluons colliding with these internal modes whose energies 
are suitable for production of the states within these wings. Thus, the 
resonance amplitude is chosen for $F(E, 0^o)$ at comparatively low energies.

Second, it follows both from experiment and from dispersion relations
that the real parts of hadronic amplitudes (and, consequently, $\Delta n$) 
become positive at very high energies as well. They are usually negative at 
intermediate energies. Their common feature is the rather high energy 
threshold above which the real parts of amplitudes become positive for all 
processes studied. 

One can expect that these general features also hold for
gluons as carriers of strong interaction forces. Then the gluonic Cherenkov 
effects can be observable in the two (low and high energy) regions.

Summarizing, the scenario, we have in mind, is as follows. Any parton, either 
belonging to a colliding nucleus or already scattered in the medium, can emit 
a gluon which traverses the nuclear medium. On its way, the gluon collides 
with some internal modes. Therefore it affects the medium as an "effective" 
wave which accounts also for the waves emitted by other scattering centers. 
Beside incoherent scattering, there are processes which can be described as 
the refraction of the initial wave along the path of the coherent wave 
(see, e.g., \cite{8}). The Cherenkov effect is the induced coherent radiation 
by a set of scattering centers placed on the way of propagation of the gluon.
The longitudinal components of their polarization vector are important. 
That is why the forward scattering amplitude plays such a crucial role in 
formation of the refractive index. At low energies its excess over 1 is 
related to the resonance peaks as dictated by the Breit-Wigner shapes of the 
amplitudes. In experiment, usual resonances are formed during the color 
neutralization process. The resonant structure can reveal itself in the virtual
states as well. However, only those gluons whose energies are within 
the left-wing resonance region of $n>1$ give rise also to the collective 
Cherenkov effect proportional to $\Delta n$. At high energies these excitations should
lead to Cherenkov jets. In both energy regions the coherent Cherenkov emission 
proceeds at angles determined by Eq. (\ref{cos}) where $n$ can depend on energy.

\section{Resonance rings.}

$\,\;\;$ {\bf Prediction 1.} According to Eq. (\ref{cos}) the ring-like 
two-dimensional distribution of hadrons similar to Cherenkov rings of photons 
can be observed in the plane perpendicular to the cone (jet) axis if $n>1$. 

{\bf Proposal 1.} Plot the one-dimensional pseudorapidity 
($\eta =-\ln \tan \theta /2$) distribution with trigger momentum as $z$-axis
neglecting the mismatch of trigger and away-side jets directions\footnote{The
mismatch asks for a higher value of the product $x_1x_2$ of the energy shares
carried by colliding partons. The probability of such events gets lower due to 
PDF decrease. Also, it reduces the cone angle. Thus its influence on $\eta $ and 
$\Delta \phi $-distributions is minimized. According to estimates in Appendix
the spread in $\Delta \phi $ can be enlarged less than by 10$^o$.}. It should 
have maximum at (\ref{cos}).

This is the best possible one-dimensional projection of the ring. 

This plot is still unavailable at RHIC. RHIC experiments \cite{5a, 6a, adar, jia} 
have shown the two-bump structure of the azimuthal angle ($\Delta \phi $)
distribution (now with 
$z$-axis chosen along the collision axis) of particles with rather low 
transverse momenta  near the away-side jets in central heavy-ion collisions. 
There is no such structure in pp-collisions. The difference has been 
attributed to "in-medium" effects. These features are clearly seen in Fig. 1 
(the upper part for pp, lower one for Au-Au). 

One easily notices the remarkable difference between particle distributions 
in the direction opposite to the trigger jet maximum positioned at 
$\Delta \phi = 0$. Both trigger and companion high-$p_T$ 
jets have been created in central Au-Au collisions at $\sqrt s=200$ GeV 
at the periphery of a nucleus. They move in opposite directions if produced in
head-on collisions of partons with equal energies. The trigger
parton immediately escapes the nucleus and, therefore, is detected as the
"in-vacuum" jet. The companion (away-side) jet traverses the whole nucleus
before it comes out. It is modified by "in-medium" effects. 

These features can be interpreted in the following way. Beside normal jet
fragmentation, the parton impact on the medium initiates emission of Cherenkov 
gluons which produce
a ring of hadrons in the plane perpendicular to the away-side jet axis.
Ring's plane is perpendicular both to the trigger momentum and to the collision 
plane in which momenta of the colliding particles and the trigger are placed.
The two-bump structure results due to the one-dimensional projection of the 
ring on the azimuthal plane. The analogous two-bump structure was shown by
Cherenkov in his earlier papers \cite{cher} (see also Fig. 1.8 in \cite{jel}).
It is clear that projection of a ring on its 
diameter in the azimuthal plane is not the best one to reveal
its properties. The Proposal 1 uses better (circular) projection of the ring. 
The shapes of two- and three-particle correlations studied at RHIC \cite{pr}
are its less direct indications although they have the ring-like structure 
themselves.

Thus we can ascribe two contributions to the away-side hadrons 
associated with the companion jet: one from jet fragmentation and 
the other from Cherenkov gluons. The hadrons from jet fragmentation are 
smoothly distributed within the phase space volume. In distinction, the 
one-dimensional distribution along the ring diameter of the away-side 
hadrons created by Cherenkov gluons must possess two peaks.

Let us note that the azimuthal angles $\Delta \phi $ in Fig. 1 
are considered as the polar angles $\theta $ in our treatment because 
the role of $z$-axis is now played by the jet axis in place of the axis of 
collisions. The ring is placed in the plane perpendicular to the 
jet axis. The two maxima in the right-hand side of Fig. 1 appear due to its 
projection on the diameter. The distance between them is exactly equal to 
the diameter of the ring.
 
 From the distance between the peaks defined in angular ($\theta =D$ in PHENIX
notation) variables in Fig. 1 one gets according to Eq. (\ref{cos}) the 
nuclear refractive index. Its value is found to be quite large $n=3$ 
compared to usual electromagnetic values for gases 
close to 1. If interpreted in terms of the Breit-Wigner resonances, as 
explained below, it results in the large density of partons in the created   
quark-gluon system with about 20 partons within the volume of a single 
nucleon \cite{6}. It agrees with its estimates from $v_2$ and hydrodynamics.
This value is also related to the energy loss of gluons 
estimated in \cite{6} as $dE/dx\approx 1$ GeV/fm. The height of the peaks 
determines the width of the ring which in its turn defines the free path 
length of Cherenkov gluons \cite{6} which happens to be long enough 
$R_f\sim 7$ fm. Thus they hadronize, probably, close to the surface of the 
initial volume.

These estimates are obtained \cite{6} as follows.
If the hadronization of gluons is a soft process then the gluon energy closely
corresponds to the energy of the produced resonance. It implies that in this
particular experiment Cherenkov gluons can be emitted only 
with energies within the lower wings of hadronic resonances. Their amplitude 
is of the Breit-Wigner shape. For a single resonance at energy $E_R$ and 
with width $\Gamma _R$ one gets
\begin{equation}
{\rm Re} n(E)=1+\frac {2J+1}{(2s_1+1)(2s_2+1)}\cdot \frac {6\mu ^3\Gamma _R\nu }
{E_R^2}\cdot \frac{E_R-E}{E[(E-E_R)^2+\Gamma _R^2/4]}.    \label{ren}
\end{equation}
Here $J$ is the spin of the resonance, $s_i$ are the spins of "incident 
particles". We have used the number of partonic scatterers $\nu $
within a single nucleon volume $4\pi /3\mu ^3$ with $\mu $ the pion mass.
For $n=3$, its estimated value $\nu \approx 20$ given above follows from 
Eq. (\ref{ren}). Let us note that the similar expressions are widely used in 
optics, in particular, for defining the variety and abundance of chemical 
elements in the sun atmosphere (e.g., see \cite{9}).

Another information can be obtained from the height of the peaks in Fig. 1.
It determines the width of the Cherenkov ring $\delta $. This is the ring 
in the plane perpendicular to the cone axis filled by evenly distributed 
within it particles over the smooth background due to jet fragmentation.
Its projection on the diameter corresponds to the particle distribution which
has a minimum at the center, increases, reaches the maximum at the internal
radius of the ring $r_i$ and then decreases to zero at its external radius 
$r_e$. For narrow rings ($\delta \ll r_i$) the height of the maximum over the
minimum is easily determined as
\begin{equation}
h_{max}=\sqrt {2r_i\delta }-\delta .     \label{max}
\end{equation}
With $h_{max} \approx 1.6 - 1.2 = 0.4$ and $r_i = 1.2$ in Fig. 1 one gets
\begin{equation}
\delta \approx 0.1.        \label{delta}
\end{equation}
The ring of Cherenkov gluons is really quite narrow. Following Proposal 1, its 
width can be directly measured by plotting the pseudorapidity distribution of 
particles with the away-side jet direction chosen as $z$-axis.

Actually, Eq. (\ref{cos}) implies that the ring is squeezed to a circle. There 
are three physical reasons which can lead to the finite width of the ring. 
First, it is the dispersion, i.e. the energy dependence of the  
refractive index. Its contribution to the width is well known
\begin{equation}
\delta _d=\int _0^{\delta _d}d\theta =\cot \theta _c\int _0^{\infty }
\frac {1}{n} \frac{dn}{dE} dE.      \label{lan}
\end{equation}
If the Breit-Wigner expression (\ref{ren}) for $n(E)$ is used, the result is
\begin{equation}
\delta _d=0.   \label{deld}
\end{equation}
It is amazing that there is no widening of the Cherenkov cone due to the 
dispersion of $n(E)$ described by the formula (\ref{ren}) with Breit-Wigner
resonances.

Second, the width of the Cherenkov ring can be due to the finite free path 
length for partons. Qualitatively, it can be estimated as the ratio of the 
parton wavelength $\lambda $ to the free path length $R_f$
\begin{equation}
\delta _f \sim \frac {\lambda }{R_f}.    \label{rf}
\end{equation}
For $\lambda \sim 2/E_R$ and $\delta _f < 0.1$ one gets the estimate for
the free path length
\begin{equation}
R_f>20/E_R \sim 4.5/\mu \sim 7\cdot 10^{-13} {\rm cm}.   \label{rfn}
\end{equation}
This appears to be quite a reasonable estimate.
The inequality sign shows that the partial width due to this particular
effect is smaller than the total width.

Finally, the width of the ring can become larger due to the processes of
resonance formation (hadronization of the gluon collective mode) and decays.
However, this can be quantified only if the Monte Carlo program for jets 
with Cherenkov gluons is elaborated.

The energy loss can be calculated using the standard formula
\begin{equation}
\frac {dE}{dx}=4\pi \alpha _S\int _{E_R-\Gamma _R}^{E_R}E\left (1-\frac {1}
{n^2(E)}\right )dE.       \label{dedx}
\end{equation}
The integration limits define the most important region discussed above. With 
expression (\ref{ren}) for $\rho $-meson one gets from (\ref{dedx})
\begin{equation}
\frac {dE}{dx} \approx 1 \; {\rm GeV/fm}.     \label{dedn}
\end{equation}
This estimate is an order of magnitude higher than the value of 0.1 GeV/fm 
obtained in the model of \cite{a} which is somewhat underestimated, 
in our opinion. It is determined by energies required 
to excite resonances. However, it is still smaller than the radiative loss.

Thus, using the RHIC data, we have estimated such parameters of the nuclear 
matter in heavy-ion collisions as its nuclear refractive index, the density 
of partons, their free path length and energy loss.

\section{Asymmetry of in-medium resonances.}

Another specific feature of low-energy Cherenkov effect is that it leads to 
the somewhat unusual particle content within the ring. 

{\bf Prediction 2.} Masses of Cherenkov states are less than in-vacuum meson 
masses. This leads to the asymmetry of decay spectra of resonances with 
increased role of low masses.

{\bf Proposal 2.} Plot the mass distribution of $\pi ^+\pi ^-$,
$\mu ^+\mu ^-$, $e^+e^-$-pairs near resonance peaks.

Apart from the ordinary Breit-Wigner shape of the cross section for 
resonance production, the dilepton mass spectrum would acquire the additional
term proportional to $\Delta n$ (that is typical for Cherenkov effects) 
at masses below the resonance peak \cite{dnec}. 
Therefore its excess (e.g., near the $\rho $-meson~\footnote{Only 
$\rho $-mesons are considered here because the most precise experimental data
are available \cite{1a} about them. To include other mesons, one should 
evaluate the corresponding sum of similar expressions. Other experimental data 
can be found in \cite{1b, 2a, 3a, 4a, 5, Muto, 6b, 7a, 8a}.}) can be described 
by the following formula
\begin{equation}
\frac{dN_{ll}}{dM}=\frac {A}{(m_{\rho }^2-M^2)^2+M^2\Gamma ^2}
\left(1+w\frac{m_{\rho }^2-M^2}{M^2 }\theta (m_{\rho}-M)\right).    \label{ll}
\end{equation}
Here $M$ is the total c.m.s. energy of two colliding objects (the dilepton
mass), $m_{\rho }$=775 MeV is the in-vacuum $\rho $-meson mass.
The first term corresponds to the Breit-Wigner cross section. According to the
optical theorem it is proportional to the imaginary part of the forward
scattering amplitude. The second term is proportional to $\Delta n$ where the 
ratio of real to imaginary parts of Breit-Wigner amplitudes is taken into 
account
\begin{equation}
\frac {{\rm Re} F(M, 0^o)}{{\rm Im} F(M, 0^o)}=\frac {m_{\rho }^2-M^2}{M\Gamma }
\label{reim}
\end{equation}
so that
\begin{equation}
\Delta n = \frac {N_s}{\Gamma }\sigma _{BW}\frac {m_{\rho }^2-M^2}{M^2},
\end{equation}
where $\sigma _{BW}$ is the Breit-Wigner cross section.

This term vanishes for $M>m_{\rho }$ in Eq. (\ref{ll}) because only positive 
$\Delta n$ lead to the Cherenkov effect. Namely it describes the distribution of
masses of Cherenkov states. In these formulas, one should take into account 
the in-medium modification of the height of the peak and its width. In 
principle, one could consider $m_{\rho }$ as a free in-medium parameter as well.
Let us rely on experimental findings that its shift in the medium is small.
Theoretical estimates would ask for some dynamics to be known. In our approach, it is
not determined. Therefore, first of all, one may just fit the parameters $A$ 
and $\Gamma $ by describing the shape of the mass spectrum at $0.75<M<0.9$ GeV 
measured in \cite{1a} and shown in Fig. 2. In this way any
strong influence of the $\phi $-meson is avoided. Let us note that 
$w$ is not used in this procedure. The values $A$=104 GeV$^3$ 
and $\Gamma =0.354$ GeV were obtained. The width of the in-medium 
peak is larger than the in-vacuum $\rho $-meson width equal to 150 MeV. 

Thus the low mass spectrum at $M<m_{\rho }$ depends only on a single 
parameter $w$ which is determined by the relative role of Cherenkov effects and 
ordinary mechanism of resonance production.  It is clearly seen from 
Eq. (\ref{ll}) that the role  of the second term in the brackets increases 
for smaller masses $M$. The excess spectrum in the mass region from 0.4 GeV 
to 0.75 GeV has been fitted by $w=0.19$. The slight downward shift 
about 40 MeV of the peak of the distribution compared with $m_{\rho }$ may
be estimated from Eq. (\ref{ll}) at these values of the parameters. This 
agrees with the above statement about small shift compared to $m_{\rho }$.
The total mass spectrum (the dashed line) and its widened Breit-Wigner 
component (the solid line) according to Eq. (\ref{ll}) with the chosen 
parameters are shown in Fig. 2. The overall description of experimental 
points seems quite satisfactory. The contribution of Cherenkov gluons (the 
excess of the dashed line over the solid one) constitutes the noticeable part 
at low masses. The formula (\ref{ll}) must be valid in the vicinity of the 
resonance peak. Thus we use it for masses larger than 0.4 GeV only.

The excess of masses larger than 0.9 GeV is ascribed to $\phi $-meson not
considered here. The estimated value $w=0.19$ corresponds to the lowest 
possible contribution of Cherenkov gluons because the fit in the region
0.75 - 0.9 GeV is most unfavorable for it. This is clearly seen in Fig. 2
where the solid line is systematically below experimental points at
0.75 - 0.8 GeV and above them at 0.8 - 0.9 GeV. 

One would expect slightly lower $p_T$ for low-mass dilepton pairs from 
coherent Cherenkov processes than for incoherent scattering at higher masses. 
Qualitatively, this conclusion is supported by experiment \cite{1a}. The
Cherenkov dominance region of masses from 400 MeV to 600 MeV below the 
$\rho $-resonance has softer $p_T$-distribution compared to the resonance 
region from 600 MeV to 900 MeV filled in by usual incoherent scattering.
More accurate statements can be obtained after the microscopic theory
of Cherenkov gluons developed. 

Whether the in-medium Cherenkov gluonic effect is as strong as shown in 
Fig. 2 can be verified by measuring the angular distribution of the lepton
pairs with different masses. The trigger-jet experiments similar to that at 
RHIC are necessary to check this prediction. One should measure the
angles between the companion jet axis and the total momentum of the lepton
pair. The Cherenkov pairs with masses between 0.4 GeV and 0.7 GeV
should tend to fill in the rings around the jet axis. The angular radius 
$\theta $ of the ring is determined by the usual condition (\ref{cos}). 

Another way to demonstrate it is to measure the average mass of lepton pairs
as a function of their polar emission angle (pseudorapidity) with the
companion jet direction chosen as $z$-axis. Some excess of low-mass pairs 
may be observed at the angle (\ref{cos}). 

In practice, these procedures can be quite complicated at comparatively low 
energies if the momenta of decay products are comparable to the transverse 
momentum of the resonance. It can be a hard task to pair leptons in reliable 
combinations. The Monte Carlo models could be of some help.

In non-trigger experiments like that of NA60 there is another obstacle. 
Everything is averaged over directions of initial  partons. Different partons 
are moving in different  directions. The angle $\theta $, measured from the
direction of their initial momenta, is the same but the total angles are 
different, correspondingly. The averaging procedure would shift the maxima 
and give rise to more smooth distribution. Nevertheless, some indications 
on the substructure with maxima at definite angles have been found at the 
same energies by CERES collaboration \cite{26}. It is not clear yet 
if it can be ascribed to Cherenkov gluons. To recover a definite maximum,
it would be better to detect a single parton jet, i.e. to have a trigger. 

The prediction of asymmetrical in-medium widening of {\bf any} resonance at its
low-mass side due to Cherenkov gluons is universal. This universality is 
definitely supported by experiment. Very clear signals of the excess on 
the low-mass sides of $\rho $, $\omega$ and $\phi$ mesons have been seen in KEK
\cite{5, Muto}. This effect for $\omega $-meson is also studied by
CBELSA/TAPS-collaborartion
\cite{7a}. Slight asymmetry of $\phi $-meson near 0.9 - 1 GeV is noticeable in
the Fig. 2 shown above but the error bars are large there. We did not try to 
fit it just to deal with as small number of parameters as possible. There are 
some indications from PHENIX at RHIC (see Fig. 6 in \cite{6b}) on this effect 
for $J/\psi $-meson. It is astonishing that this effect has been observed in
a wide interval of initial energies. The relative share of Cherenkov effects,
described by the parameter $w$ above, can depend on energy.

To conclude, the universal asymmetry of in-medium mesons with an excess over
the usual Breit-Wigner form at low masses is predicted as a signature of
Cherenkov gluons produced with energies which fit the left wings of resonances.

Let us stress that we do not require $\rho $-mesons or other resonances 
pre-exist in the medium but imply that they are the modes of its excitation 
formed during the hadronization process of partons. 
The Cherenkov gluon emission is a collective response of the quark-gluon medium 
to impinging partons related to its preconfinement and hadronization properties. 
It is defined by energy behaviour of the second term in Eq.~(\ref{ren}). 

For the sake of simplicity, Eqs. (\ref{delt}) and (\ref{ren}) valid at small 
$\Delta n_R$ typical for gases are used here. The value $n=3$ corresponds to a 
dense liquid. Therefore, one must use~\cite{9}
\begin{equation}
\frac {n^2-1}{n^2+2}=\frac {m_{\pi }^3\nu \alpha }{4\pi }=\sum _R
\frac {2J_R+1}{(2s_1^R+1)(2s_2^R+1)}\cdot \frac {4m_{\pi }^3\Gamma _R \nu }
{EE_R^2}\cdot \frac {E_R-E}{(E-E_R)^2+\Gamma _R^2/4},   \label{liq}
\end{equation}
where $\alpha $ denotes the colour polarizability of the colour-neutral 
medium. The value 
$\nu $ obtained from this expression is almost twice lower than given above.
It does not change the qualitative conclusions about the dense medium 
(for more details see \cite{6, 7}).

\section{High-energy rings.}

At much higher energies one can expect better alignement of the momenta of
initial partons. This would favour the direct 
observation of emitted by them rings in non-trigger experiments. The first 
cosmic ray event \cite{23} with ring structure gives some hope that at LHC 
energies the initial partons are really more aligned and this effect can be 
found. The possible additional signature would be the enlarged transverse 
momenta of particles within the ring.

{\bf Prediction 3}. The very high energy forward moving partons can emit 
high energy Cherenkov gluons producing jets.

{\bf Proposal 3}. Plot the pseudorapidity distribution of dense groups of
particles in individual events (now again with collision axis chosen as 
$z$-axis) and look for maxima at angles determined by Eq. (\ref{cos}).

Gluons with such energy are not abundant at RHIC but they will become available 
at LHC. Namely such gluons were discussed in \cite{1, 2} in connection with the 
cosmic ray event at energy $10^{16}$ eV (in the target rest system $E_t$) with 
the ring-like structure first observed \cite{23}. This energy just corresponds 
to LHC energies. The partons emitting such gluons move with high energy in 
the forward direction. 
With ${\rm Re} F(E_t)$ fitted to experimental data and dispersion relation 
predictions at high energies one can expect (see \cite{1, 2}) that the excess
of $n$ over 1 behaves as
\begin{equation}
\Delta n_R(E_t)\approx \frac {a\nu_h}{E_t}\theta (E_t-E_{th}).   \label{nh}
\end{equation}
Here, $a\approx 2\cdot 10^{-3}$ GeV is a parameter of ${\rm Re} F(E_t)$ obtained
from experiment (with dispersion relations used) and $\nu _h$ is the parton 
density for high energy region. It can differ from $\nu $ used at low energies. 
$\Delta n_R(E_t)$ is small and decreases with energy for constant $\nu _h$. 
It would imply that the medium reminds a gas but not a liquid for very high 
energy gluons, i.e. it becomes more transparent.

The angles of the cone emission in c.m.s. of LHC experiments must be very large 
nevertheless (first estimates in \cite{2} are 60$^o$ - 70$^o$), i.e. 
the peaks can be seen in the pionization region at central pseudorapidities. In 
more detail it is discussed in \cite{1, 2, 8}. In this region the background 
is large, and some methods to separate the 
particles in the cone from the background were proposed in \cite{ag, dst}. 

The main difference between the trigger experiments at RHIC and this 
nontrigger experiment is in the treatment of the rest system of the medium. 
The influence of the medium motion on cone angles was considered in \cite 
{ssm}. It is important because all the above formulas are valid for emission 
in the rest system of the medium. 

At RHIC, the 90$^o$ trigger jet defines 
the direction of the away-side jet. Because of position of the trigger 
perpendicular to the collision axis of initial ions, the accompanying 
partons (particles) feel the medium at rest on the average in the c.m.s.
The similar trigger experiments are possible at LHC.
It is important to measure the cone angles for different angular 
positions of the trigger to have an access to different coordinate systems
where the target is at rest on the average and to register the medium motion. 

In nontrigger experiments, dealing with forward 
moving high energy partons inside of one of the colliding ions, the rest 
system of the medium is the rest system of another colliding ion. Therefore 
the cone angle should be calculated in that system and then transformed to the 
c.m.s. That is why these angles are so large even at small values of the 
refractivity index for high energy gluons. The low energy Cherenkov gluons 
emitted at rather large angles in the rest system of this nucleus are hard 
to observe because at LHC they fly backward inside the accelerator pipe 
(with $\vert y\vert \approx 8$, i.e. close to 180$^o$ in c.m.s.).

The idea about Cherenkov gluons was first used to interpret the cosmic ray 
event at energy 10$^{16}$ eV \cite{1, 23} where two rings more densely 
populated by particles than their surroundings were noticed. It is demonstrated
in Fig. 3 where the number of produced particles is plotted as a function of 
the distance from the collision point. It clearly shows two maxima.  They
correspond to two maxima on the pseudorapidity scale which would arise due to 
two rings produced. Again, the medium rest system coincides with the target 
rest system. 

This event has been registered in the detector with 
nuclear and X-ray emulsions during the balloon flight at the altitude about
30 km. The most indefinite characteristics of the event is the height $H$
over the detector at which the interaction took place. However, it can be
estimated if one assumes that the two rings observed with radii $r_1=1.75$ cm
and $r_2=5$ cm are produced, correspondingly, by forward and backward moving 
(in c.m.s.) partons. Using the transformation of angles from target ($t$) 
to c.m.s. ($c$) 
\begin{equation}
\tan \frac {\theta _c}{2}\approx \gamma \theta _t   \;\;\;\; and \;\;\;\;
\frac{\theta _{1t}}{\theta _{2t}}=\frac{r_1}{r_2}                      \label{tan}
\end{equation}
and assuming the symmetry of rings $\theta _{2c}=\pi - \theta _{1c}$, one gets 
$\theta _{1c}\approx 61^o$ and $\gamma \approx 2.3\cdot 10^3$.
The angle in the target rest system is $\theta _{1t} \approx 2.6\cdot 10^{-4}$ 
and the height $H=r_1/\theta _{1t} \approx 68$ m. It corresponds rather 
well to the experimental estimates obtained by three different methods
\cite{kon}. The most reliable of them give values ranging from 50 m to 100 m. 
Even though the observed angles were quite small in the target rest system, 
at LHC they would correspond to large c.m.s. angles about 
60$^o$ - 70$^o$. The peaks in the angular distribution of jets over the 
background at these angles would be observable \cite {ag, dst}. 

The substructure similar to that in Fig. 3 was also found in PbPb-interactions
at 158 GeV. The content of peaks of both cosmic ray and PbPb-events
was revealed by the wavelet analysis \cite{adk, dikk} of the 
particle distribution in the plane perpendicular to the axis of collision. 
Fig. 4 shows the dark regions with large wavelet coefficients corresponding to
dense groups of particles in a PbPb-event. There is a tendency for these groups 
to fill in the two rings. We ascribe them to Cherenkov gluonic rings. 

The multiplicities at LHC will be much higher. Therefore the wavelet analysis 
of very high multiplicity events can be more effective and should be used for
search of dfferent patterns like jets, rings, fractality, elliptic flow, higher 
Fourier coefficients, i.e., in general, for event-by-event studies of the 
three-dimensional phase space structure.
Fig. 4. The two-dimensional ($\eta - \phi $) distribution of particles in
a PbPb-event at 158 GeV as revealed by the wavelet analysis. \\

The problem of the finiteness of the nuclear target was discussed in \cite{2}. 
The criterium for the target size $L$ to be considered as infinite is
\begin{equation}
E_tL\Delta n \gg 1  \;\;\;\; {\rm or} \;\;\;\; L \gg \frac {1}{E_t\Delta n}.
\end{equation}
It is well fulfilled in the RHIC trigger experiment. Here $\Delta n \approx 2$
and $E_t$ is approximately given by the energy of emitted Cherenkov states.
It asks for some care at small $\Delta n$, however. 

\section{Discussion and conclusions.}

One of the most intriguing problems is that the RHIC and cosmic ray data
were fitted with very different values of the refractive index equal to
3 and close to 1, correspondingly. This could be interpreted as due to the
difference in values of $x$ (the parton share of energy) and $Q^2$ (the
transverse momenta). It is well known that the region of large $x$ and $Q^2$
corresponds to the dilute partonic system. At low $x$ and $Q^2$ the density
of partons is much higher.

As clearly stated in the PDG report \cite{pdg}, "the kinematical ranges of
fixed-target and collider experiments are complemetary, which enables the
determination of PDF's over a wide range in $x$ and $Q^2$".

At RHIC one deals with rather low $x$ and $Q^2$. One would 
expect the large density of partons in this region and, therefore, high $n$.
It is interesting to note that the two-bump structure disappears in RHIC data 
at higher $p_t$ where the parton density must get lower. It corresponds to 
smaller $n$ and $\theta $, i.e. bumps merge in the main away-side peak.
  
In the cosmic ray event one observes effect due to leading partons with large 
$x$. Also, the experimentalists pointed out that the transverse momenta in this 
event are somewhat enlarged \cite{kon}. In this region one would expect for low
parton density and small $n$. 

Thus the same medium can be seen as a liquid or a gas depending 
on the parton energy and transferred momenta. This 
statement can be experimentally verified by using triggers positioned at 
different angles to the collision axis and considering different transverse 
momenta. In that way, the hadronic Cherenkov effect can be used as a tool to 
scan ($1/x, Q^2$)-plane and plot on it the parton densities corresponding
to its different regions.  

The experimental separation of "mismatched" events with direction of the
away-side jet not opposite to the trigger jet can also help in scanning
this plane.

The interesting theoretical domain not yet explored is the longitudinal
nature of the polarization field responsible for the Cherenkov effect.
One can speculate that the non-perturbative string forces become more
important than pQCD parton interactions.

To conclude, the predictions about Cherenkov gluons and proposals for their 
verification are presented. The ring-like structure of events is discussed. 
The properties of the nuclear medium are determined from the data about
the rings. The universal asymmetry of in-medium resonances is described as 
a signature of Cherenkov gluons. The low- and high-energy effects for different
trigger positions and events with away-side jets unaligned to the trigger
jet can be used to scan regions with various parton densities.\\

{\bf Acknowledgments}

\noindent I am grateful to A.V. Leonidov for comments and to V.A. Nechitailo 
for collaboration. This work has been supported in part by the RFBR grants 
06-02-16864, 06-02-17051.\\

\section{Appendix.}

\begin{center} 

       {\large   The mismatch of jets in trigger experiments.}               \\

\end{center} 


Let us first consider the experiment where the jet trigger is positioned 
at $\pi $/2 to the collision axis and registers jets with energy $p_{tr}$.
These jets are initiated by scattering of two partons with energies $x_1p$
and $x_2p$. If $x_1=x_2=x$, this is the scattering at $\pi $/2 with two jets of
equal ($p_{tr}=xp$) and opposite momenta created. There is no mismatch in 
their emission angles. If $x_1$ differs from $x_2$, such a mismatch appears.
For a fixed value of the trigger jet momentum $p_{tr}$, only collisions with
a definite relation between the momenta of colliding partons $x_ip$ can
initiate such a process.

The energy-momentum conservation requires that 
\begin{equation} 
x_2=\frac {x_1p_{tr}}{2x_1p-p_{tr}},
\end{equation} 
i.e. a special asymmetry of initial parton momenta is required for this 
process to proceed. No mismatch case corresponds to $x_1=x_2=p_{tr}/p$.
As an example of a mismatch, considering $x_1=2p_{tr}/p$ one gets $x_2=2p_{tr}/3p$.
The momentum of the away-side jet $p_a$ is 
\begin{equation} 
p_a=x_1p-\frac {(x_1p-p_{tr})p_{tr}}{2x_1p-p_{tr}}.
\end{equation} 
The angle of its emission determines the angular mismatch
\begin{equation}
\sin \theta _a=p_{tr}/p_a.
\end{equation}

For a general case of the trigger positioned at the angle $\theta $ to the
collision axis one gets
\begin{equation}
x_2=\frac {x_1p_{tr}(1-\cos \theta)}{2x_1p-p_{tr}(1+\cos \theta)},
\end{equation}
\begin{equation} 
p_a=x_1p-\frac {(x_1p-p_{tr})p_{tr}(1+\cos \theta)}{2x_1p-p_{tr}(1+\cos \theta)},
\end{equation} 
\begin{equation}
\sin \theta _a=p_{tr}\sin \theta /p_a.
\end{equation}
The probability of mismatched jets is lower by a factor $f(x_1)f(x_2)$ where
$f$ is pdf.

To conclude, the jet trigger at fixed energy and angle chooses a well defined
set of events with mismatched energies of initial colliding partons. It can
be used to measure the $x$-dependence of pdfs.

The impact of the mismatched jets on the azimuthal projection of Cherenkov 
rings can be estimated. In no mismatch case the maximum angle $\Delta \phi _0$
is equal to the cone angle $\theta _c=1/n$. Using the ratio of the radius of
the ring $r_0$ to the distance from the collision vertex $h_0$ one gets
\begin{equation}
\tan \Delta \phi _0=r_0/h_0=\tan \theta _c.
\end{equation}
The ring around the mismatched away-side jet projected on the azimuthal plane
is spread up to
\begin{equation}
\tan \Delta \phi _m =\frac {\tan \Delta \phi _0}{\sin \theta _a}=
\frac {p_a \tan \Delta \phi _0}{p_{tr}\sin \theta }.
\end{equation}
The mismatch widens the $\Delta \phi $-distribution of ring projections.
These formulas are valid for $\theta _a\geq \theta _c$, i.e. for comparatively 
small mismatch. For strongly different $x_1$ and $x_2$ the part of the cone 
can even enter the opposite hemisphere (if $\theta _c\geq \theta _a$) .
However the diminished probability of the mismatched jets and the decreased
intensity of the projection due to the turned by the angle $\theta _a$ ring
should reduce the influence of this shift.

Let us consider two examples of mismatch for the $\pi /2$-trigger.

For $x_1=2p_{tr}/p$ one gets $x_2=2p_{tr}/3p$, $p_a=5p_{tr}/3, \sin \theta _a=
0.6, f(x_1)f(x_2)=0.75^{1.3}\approx 0.688$ for $f(x)\propto x^{-1.3}$.
If $\Delta \phi _0=\pi /4$, widened $\Delta \phi _m=1.03$, i.e. 59$^o$ instead
$45^o$. 
If $\Delta \phi _0=\pi /3$, widened $\Delta \phi _m=1.237$, i.e. 71$^o$ instead
$60^o$. 

For $x_1=3p_{tr}/p$ one gets $x_2=0.6p_{tr}/p$, $p_a=2.6p_{tr}, \sin \theta _a=
5/13, f(x_1)f(x_2)=(5/9)^{1.3}\approx 0.466$ for $f(x)\propto x^{-1.3}$.
If $\Delta \phi _0=\pi /4$, widened $\Delta \phi _m=1.2$, i.e. 69$^o$ instead
$45^o$. 
If $\Delta \phi _0=\pi /3$, widened $\Delta \phi _m=1.352$, i.e. 77.5$^o$ instead
$60^o$. 

For RHIC trigger with $p_{tr}=5$ GeV, the mismatched initial partons in the
above examples have energies $p_1=10$ GeV, $p_2=3.33$ GeV and $p_1=15$ GeV, 
$p_2=3$ GeV, correspondingly. 

If the angular shifts are boldly weighted by corresponding $f(x_1)f(x_2)$,
one gets the shifts from 45$^o$ to 54.7$^o$ and from 60$^o$ to 67.3$^o$.

Thus, the account of mismatched jets can somewhat reduce the estimate of $n$
(and, therefore, the parton density) obtained without it but not very strongly
according to the above values of angular shifts. \\

{\bf Figure captions.}\\

Fig. 1. The $\Delta \phi $-distribution of particles produced by trigger and 
companion jets at RHIC \cite{5a} shows two peaks in $pp$ and three 
peaks in AuAu-collisions (two of them are on the away side). \\

Fig. 2. Excess dilepton mass spectrum in semi-central In-In collisions 
at 158 AGeV (NA60 data are shown by dots) compared to the in-medium 
$\rho $-meson peak with additional Cherenkov effect (the dashed line).\\

Fig. 3. The dependence of the number of produced hadrons on the distance from 
the collision point in the cosmic ray event.\\

Fig. 4. The two-dimensional ($\eta - \phi $) distribution of particles in
a PbPb-event at 158 GeV as revealed by the wavelet analysis. \\

\end{document}